\documentclass{aa}  

\usepackage{graphicx}
\usepackage{txfonts}

\begin{document}

   \title{Cloud-cloud collision and star formation in G323.18+0.15}


   \author{Yingxiu Ma
          \inst{1,2}
          \and
          Jianjun Zhou\inst{1,3,4}
          \and
          Jarken Esimbek\inst{1,3,4}
          \and
          Willem Baan\inst{1,5}
          \and
          Dalei Li\inst{1,3,4}
          \and
          Yuxin He\inst{1,3,4}
          \and
          Xindi Tang\inst{1,3,4} 
          \and
          Weiguang Ji\inst{1,3,4}
          \and
          Dongdong Zhou\inst{1,3,4}
          \and
          Gang Wu\inst{6} 
          \and
          Ye Xu\inst{7} 
          }
             
   \institute{XingJiang Astronomical Observatory, Chinese Academy of Sciences,
             Urumqi 830011, PR China\\
              \email{mayingxiu@xao.ac.cn; zhoujj@xao.ac.cn}
         \and
             University of Chinese Academy of Sciences, Beijing 100049, PR China
         \and
             Key Laboratory of Radio Astronomy, Chinese Academy of Sciences, Urumqi 830011, PR China
         \and
             Xinjiang Key Laboratory of Radio Astrophysics, Urumqi 830011, PR China
        \and
             Netherlands Institute for Radio Astronomy, ASTRON, 7991 PD Dwingeloo, The Netherlands
         \and
            Max-Planck-Institut f\"{u}r Radioastronomie, Auf dem H\"{u}gel 69, D-53121 Bonn, Germany
         \and
            Purple Mountain Observatory, Chinese Academy of Sciences, Nanjing 210008, PR China
             }             



  \abstract
  {We studied the cloud-cloud collision candidate \object{G323.18+0.15} based on signatures of induced filaments, clumps, and star formation. We used archival molecular spectrum line data from the SEDIGISM $^{13}$CO\,($J$\,=\,2--1) survey, from the Mopra southern Galactic plane CO survey, and  infrared to radio data from the GLIMPSE, MIPS, Hi-GAL, and SGPS surveys. Our new result shows that the \object{G323.18+0.15} complex is 3.55\,kpc away from us and consists of three cloud components, G323.18a, G323.18b, and G323.18c. G323.18b shows a perfect U-shape structure, which can be fully complemented by G323.18a, suggesting a collision between G323.18a 
  and the combined G323.18bc filamentary structure. 
  One dense compressed layer (filament) is formed at the bottom of G323.18b, where we detect a greatly increased velocity dispersion. 
  The bridge with an intermediate velocity in a position-velocity diagram appears between G323.18a and G323.18b, which corresponds to
  the compressed layer. 
  G323.18a plus G323.18b as a whole are probably not gravitationally bound. This indicates that high-mass star formation in 
  the compressed layer may have been caused by an accidental event. 
  The column density in the compressed layer of about $1.36 \times 10^{22}$\,cm$^{-2}$ and most of the dense clumps and 
  high-mass stars are located there. 
  The average surface density of class\,I and class\,II young stellar objects (YSOs) inside the \object{G323.18+0.15} complex is much higher than the density 
  in the surroundings. 
  The timescale of the collision between G323.18a and G323.18b is $1.59$\,Myr. This is longer than the typical lifetime of class\,I YSOs 
  and is comparable to the lifetime of class\,II YSOs.
}


 \maketitle
%

\section{Introduction}

While high-mass star formation is currently poorly understood, 
two main mechanisms for high-mass star formation have been proposed: the competitive accretion in a massive self-gravitating 
system \citep{Bonnell2001}, and the monolithic collapse of a dense cloud \citep{McKee2002, Krumholz2009}. Both mechanisms have their 
problems \citep{Fukui2020}. 
For example, the first cannot explain the formation of single isolated O-stars with a lower system mass \citep{Ascenso2018}. 
The second requires that the compact cloud or clump produces a column density $\ge$1\,g\,cm$^{-2}$ \citep{Krumholz2009,Krumholz2012}, 
but it is not clear how such a high-density molecular cloud is formed. 
If the timescale of forming a high-density cloud is comparable to the free-fall timescale, low-mass stars will form first when the cloud 
density is high enough, and they may prevent the formation of high-mass stars \citep{Fukui2020}. 
Therefore, high-density clouds have to be formed in a time much shorter than the free-fall time. 
Rapid external compression provides an alternative scenario for creating such high-density clouds in a very short time \citep{Zinnecker2007}.

Cloud-cloud collisions (CCCs) can result in the rapid accumulation of cloud mass into a small volume and form massive star-forming 
clumps within the shock interface  \citep{Inoue2013, Takahira2014, Takahira2018, Fukui2020}. 
This provides a promising mechanism that can explain isolated or asymmetric O-star formation and the production of dense 
and massive clumps forming high-mass stars \citep{Fukui2020}. 
Previous observations have identified more than 50 high-mass star formation regions with collisional features, but more studies are needed 
to fully understand the detailed properties of colliding clouds and the high-mass star formation triggered by the CCCs. 

The \object{G323.18+0.15} complex (see Fig.\ref{fig1}) has been studied as component F in the G323 region 
using data from the Mopra 22m single-dish telescope survey of the southern Galactic plane \citep{Burton2013}. 
The results suggest that the \object{G323.18+0.15} complex has a system velocity of -65\,km\,s$^{-1}$ and a distance of 4.8\,kpc. 
Its gas mass and column density are about 1.9\,$\times$\,10$^4$\,M$_\odot$ and 9.4\,$\times$\,10$^{21}$\,cm$^{-2}$. However, the kinematics of the region has not been studied in detail.

In this paper, we study the kinematics of the \object{G323.18+0.15} complex. We suggest that it may be a good candidate of 
a CCC complex. 
We describe the data we used in Sect.\,2 and present the physical parameters of the cloud region in Sect.\,3. 
Sect.\,4 discusses the evidence of CCCs and the possible effects from an associated H$\rm_{II}$ region. 
Finally, Sect.\,5 gives the conclusion of our work.


\section{Archive data}
\renewcommand{\thefootnote}{\fnsymbol{footnote}}
\subsection{$^{12}$CO and $^{13}$CO data}
For a kinematic analysis, data of the Structure, Excitation and Dynamics of the Inner Galactic 
Interstellar Medium (SEDIGISM\footnote{This publication is mainly based on data acquired with the Atacama Pathfinder EXperiment (APEX) under programs 092.F-9315(A) and 193.C-0584(A). APEX is a collaboration between the Max-Planck-Institut für Radioastronomie, the European Southern Observatory, and the Onsala Space Observatory. The processed data products are available from the SEDIGISM survey database located at https://sedigism.mpifr-bonn.mpg.de/index.html, which was constructed by James Urquhart and is hosted by the Max-Planck-Institut für Radioastronomy.}) survey have been used \citep{Schuller2021}. This survey mapped 84\,deg$^2$ of the Galactic plane 
($-60^{\circ} <l < 31^{\circ}$, $ \left| b \right| < 0.5^{\circ}$) with the Atacama Pathfinder Experiment (APEX) telescope in several molecular transitions, including $^{13}$CO\,($J$\,=\,2--1) and C$^{18}$O\,($J$\,=\,2--1). 
The angular resolution of SEDIGISM is $\sim$\,$30''$ , and the 1\,$\sigma$ sensitivity is about 0.8--1.0\,K for 
a 0.25\,km\,s$^{-1}$ channel width. 
The $^{12}$CO\,($J$\,=\,1--0) data from the Mopra 22m single-dish telescope survey of the southern Galactic plane \citep{Burton2013} 
has also been used, which has a spatial resolution of $35^ {''}$ and the 1\,$\sigma$ sensitivity of 1.5 K per 0.1\,km\,s$^{-1}$ velocity channel.

\subsection{IR and radio data}
We used the image data from the Galactic Legacy Infrared Mid-Plane Survey Extraordinaire (GLIMPSE) \citep{Churchwell2009} to derive the mid-infrared emission at 3.6, 4.5, 5.8, and 8\,$\mu$m. The 5\,$\sigma$ sensitivities at the four bands are 0.2, 0.2, 0.4, and 0.4\,mJy, respectively. The corresponding angular resolutions are between $1.5''$ and $1.9''$ \citep{Fazio2004}. We also used the image data from the Multiband Infrared Photometer for Spitzer MIPS Galactic Plane Survey (MIPSGAL) \citep{Carey2009} at 24 and 70\,$\mu$m with a 5\,$\sigma$ sensitivity of 1.7\,mJy and with corresponding resolutions of  $6''$ and $18''$, respectively \citep{Rieke2004}.
 
We used Herschel Infrared Galactic Plane Survey (Hi-GAL , \citet{Molinari2010}) data to derive the dust temperature and column density distributions. Hi-GAL is a key project of the 3.5\,m orbiting Herschel telescope as it mapped the entire Galactic plane within $ \left| b \right| < 1^{\circ}$ at five bands. The angular resolutions of these Herschel maps are approximately $10.2''$, $13.5''$, $18.1''$, $25.0''$, and $36.4''$ at 70, 160, 
250, 350, and 500\,$\mu$m, respectively. 

The 1.4 GHz radio continuum emission data from the Southern Galactic Plane Survey (SGPS, \citet{Haverkorn2006}) were used to trace H$\rm_{II}$ regions, which was observed with the Australia Telescope Compact Array and the Parkes 64\,m single-dish telescope. The survey spans a Galactic longitude of 253$^{\circ} < l < 358^{\circ}$ and a latitude of $ \left| b \right| < 1.5^{\circ}$ at a resolution of $100''$ and a sensitivity below 1\,mJy\,beam$^{-1}$.

\subsection{Catalogs}
The APEX Telescope Large Area Survey of the Galaxy (ATLASGAL) dense clump catalog \citep{Urquhart2018} and the Herschel Hi-GAL clump catalog \citep{Elia2017} were used to trace dense clumps. The GLIMPSE Point-Source catalog (GPSC) is used to trace the young stellar objects (YSOs).

   \begin{figure*}
   \centering
   \includegraphics[trim={1cm 0cm 3cm 1cm},clip,width=10cm,angle=90]{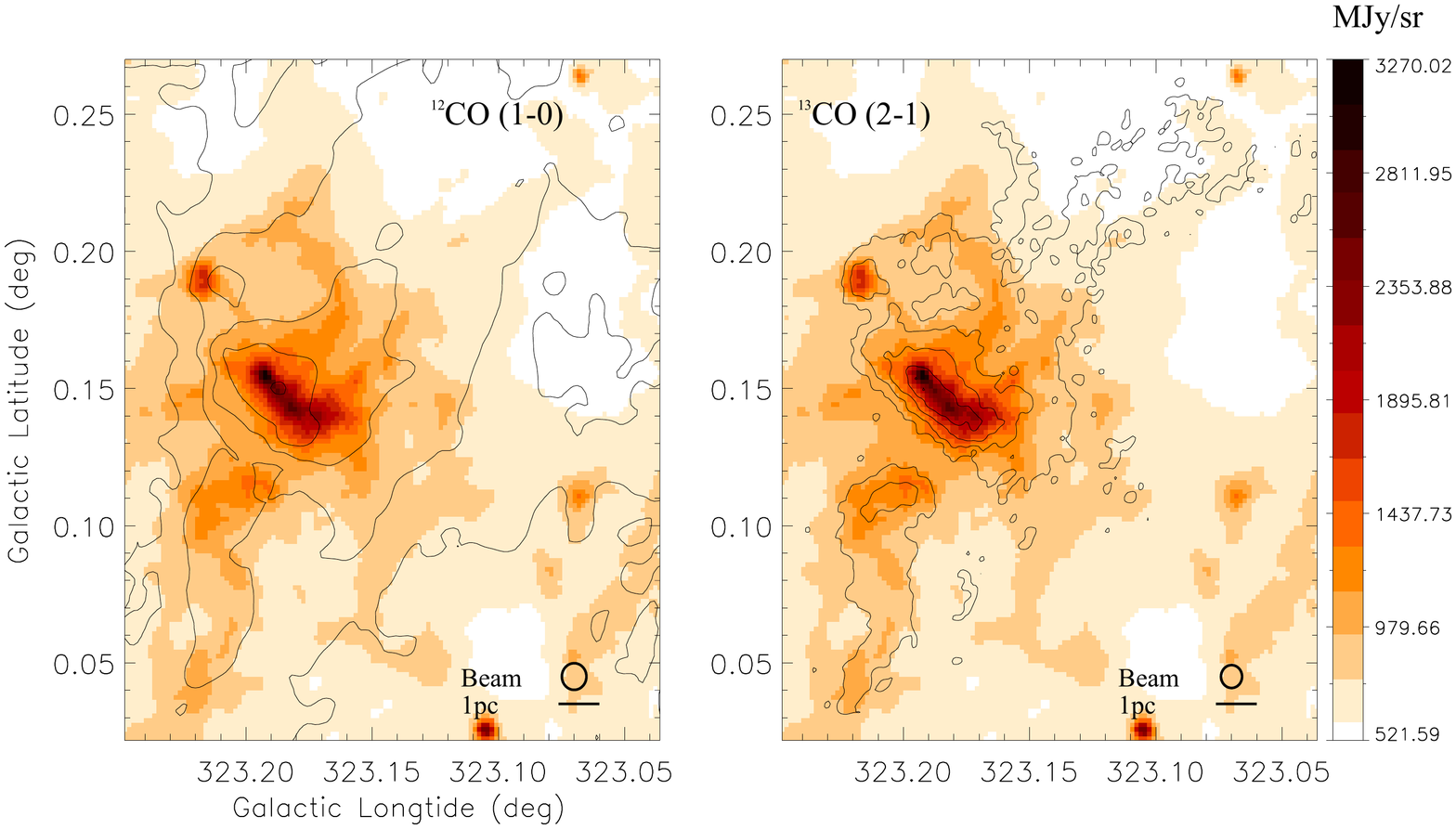}
      \caption{\object{G323.18+0.15} complex region imaged by Herschel at 250\,$\mu$m. The black contours represent the $^{12}$CO\,($J$\,=\,1--0) (left) 
      emission and $^{13}$CO\,($J$\,=\,2--1) (right) emission integrated from -69 to -61\,km\,s$^{-1}$. The contour levels start from 6 and 
      2\,K\,km\,s$^{-1}$ and increase in steps of 8 and 11\,K\,km\,s$^{-1}$, respectively.
              }
         \label{fig1}
   \end{figure*}

   \begin{figure*}
   \centering
   \includegraphics[trim={0cm 0cm 0cm 3.5cm},clip,width=18cm,angle=0]{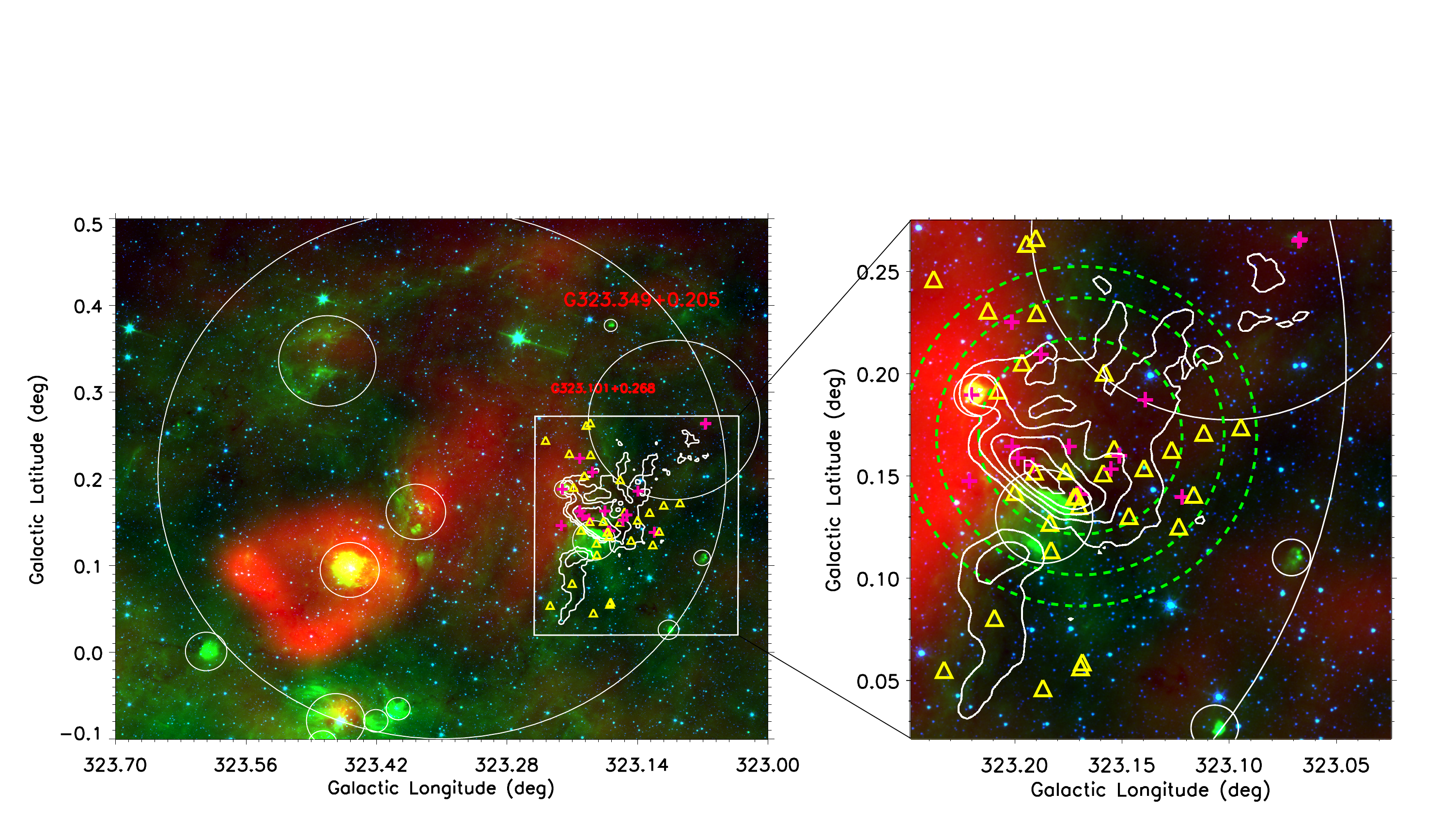}
      \caption{ Three-color map of the \object{G323.18+0.15} complex (right) and nearby H$\rm_{II}$ regions (left). 
  Left: Red, green, and blue background shows 21\,cm, $8\,\mu$m, and $4.5\,\mu$m emission, respectively. 
  The white circles are the H$\rm_{II}$ regions identified by \citet{Anderson2014}. 
  The white rectangular box is the research region, a zoom of which is shown in the right panel. 
  The white contours are the integral intensity of the $^{13}$CO emission from -69 to $-61$\,km\,s$^{-1}$ in the \object{G323.18+0.15} complex. 
  The white crosses and yellow triangles are class I and II YSOs, respectively. Right: Same as the left panel, but the  dashed green circles are the area for which the surface density was calculated (see sect.\,4.2).}
         \label{fig2}
   \end{figure*}
   
\begin{figure*}
\centering
\includegraphics[trim={1cm 0cm 2cm 2cm},clip,width=12cm,angle=90]{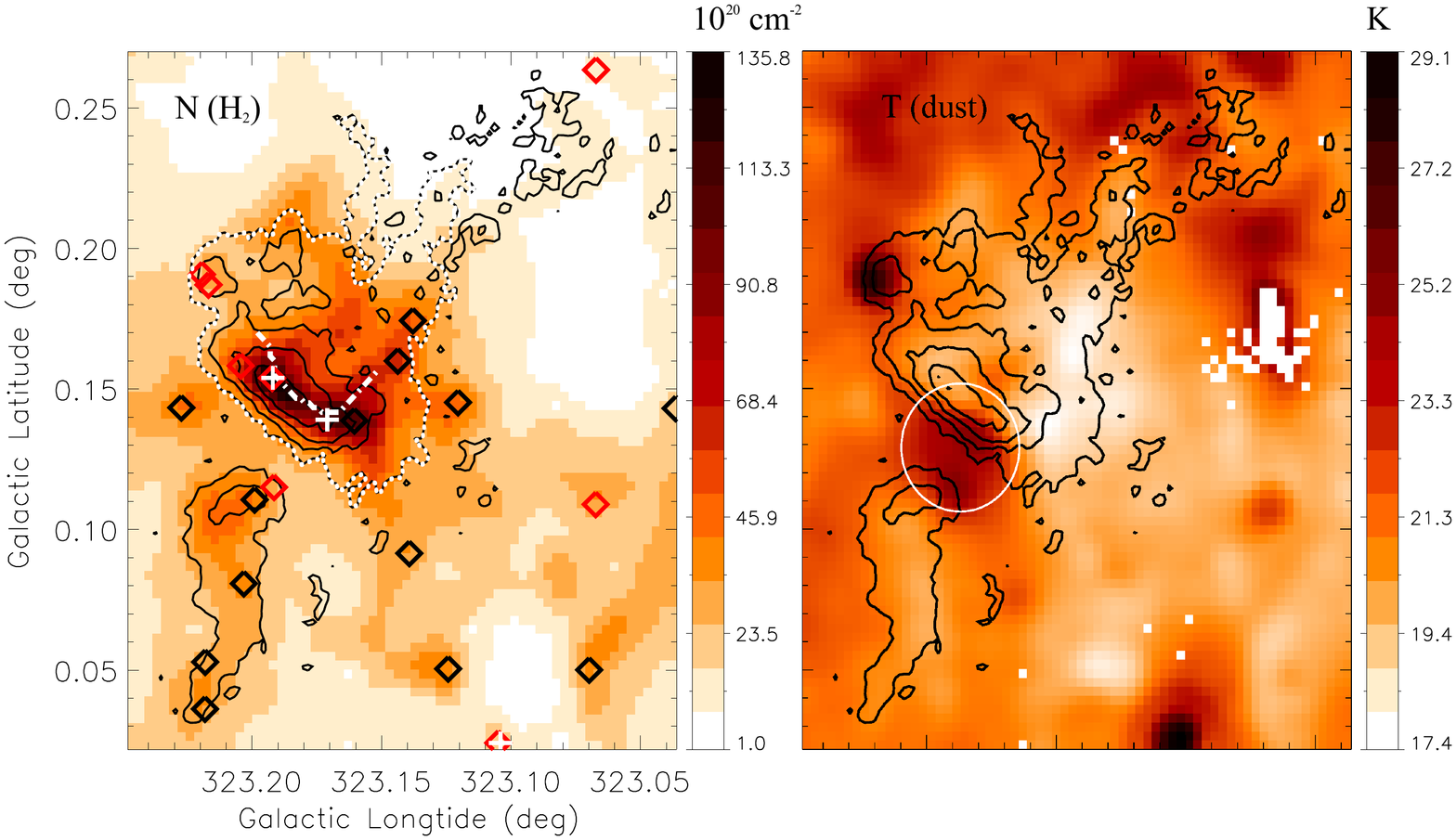}
\caption{ Column density (left) and temperature (right) distributions within the \object{G323.18+0.15} complex. 
The black contours indicate the $^{13}$CO\,($J$\,=\,2--1) integrated intensity in areas where $T_{\rm mb}$ is higher than 3\,$\sigma$ (3\,K), 
The contour levels start from 2\,K\,km\,s$^{-1}$ and increase in steps of 11\,K\,km\,s$^{-1}$. 
The velocity interval of the integration is -69 to -61\,km\,s$^{-1}$. 
The region surrounded by the dotted white line is used for the calculations in sect.\,4.1. 
The dot-dashed white line is the skeleton of filament G323.179+0.149 identified by \citet{Li2016}. 
The black and red diamonds are prestellar and protostellar clumps from the Herschel HIGAL clump catalog, and the white crosses 
are ATLASGAL 870\,$\mu$m clumps (see table.\,\ref{table:1} for details).
The white circle in the right panel is the H$\rm_{II}$ region G323.187+00.129 identified by \citet{Anderson2014}.}
\label{fig3}
\end{figure*}
   
\section{Results}

    \subsection{\object{G323.18+0.15} complex}

The distributions of the integrated $^{12}$CO\,(1--0) and $^{13}$CO(2--1) emission of the \object{G323.18+0.15} complex are shown on 
a background of the 250\,$\mu$m dust emission in Fig.\,\ref{fig1}. 
The distributions of $^{12}$CO\,($J$\,=\,1--0) and $^{13}$CO\,($J$\,=\,2--1) agree well with that of the 250\,$\mu$m dust emission. 
The $^{13}$CO\,($J$\,=\,2--1) map of the \object{G323.18+0.15} complex shows separate northwestern and southeastern components, 
while the $^{12}$CO\,($J$\,=\,1--0) map shows that these components are connected to each other. Fig.\,\ref{fig2} presents more observation results of the \object{G323.18+0.15} complex (right panel) and nearby H$\rm_{II}$ regions (left panel) at different wavelengths. White circles denote the H$\rm_{II}$ regions identified by \citet{Anderson2014}, and red background traces the 21cm continuum emission. The H$\rm_{II}$ region G323.187+00.129 is right in between the northwestern and southeastern components of the \object{G323.18+0.15} complex. The 8\,$\mu$m emission appears in all H$\rm_{II}$ regions. In the region we studied, class I (pink crosses) and class II (yellow triangles) YSOs mainly lie on the \object{G323.18+0.15} complex or nearby it.

Considering the high resolution and good data quality of $^{13}$CO\,($J$\,=\,2--1), we mainly used $^{13}$CO\,($J$\,=\,2--1) in the following analysis. Using the system velocity -65.75\,km\,s$^{-1}$ derived from the averaged $^{13}$CO\,($J$\,=\,2--1) spectrum of the \object{G323.18+0.15} 
complex and the new kinematic distance estimator\footnote{http://bessel.vlbi-astrometry.org/node/378} developed by \citet{Reid2019}, 
we obtain a distance of about $3.55 \pm 0.45$\,kpc. 
This is smaller than the earlier distance of 4.8\,kpc obtained by \citet{Burton2013}. 
Because the latest distance estimator is based on the much improved rotation curve of the Galaxy, we use the newer 
distance of \object{G323.18+0.15} complex in this work.

    \subsection{Column density and dust temperature}

First, we removed the background emission from the image data at different bands separately. 
Then we used the kernels provided by \citet{Aniano2011} to convolve all images at 70, 160, 250, and 350\,$\mu$m to the angular 
resolution $36.4''$ at 500\,$\mu$m. 
Finally, we regridded all images at five bands to the same pixel size ($11.5''$), and performed a spectral energy distribution fit pixel by pixel with the PYTHON package  HIGAL-sed-fitter\footnote{http://hi-gal-sed-fitter.readthedocs.org}. 
The code is based on a modified blackbody model,
      \begin{equation}
      I_\nu = B_\nu({1 - e^{-\tau_\nu}})\,,
      \end{equation}
       where the Planck function $B_\nu$ is modified by the optical depth \citep{Kauffmann2008},
\begin{equation}
\tau_\nu = \mu_{\rm H_2} m{\rm _H} K_\nu{N_{\rm H_2}} / R_{\rm gd},
\end{equation}
where $\mu_{\rm H_2}$\,=\,2.8 is the mean molecular weight, 
$m_{\rm H}$ is the mass of a hydrogen atom, 
the gas to-dust ratio is $R_{\rm gd}$\,=\,100, 
and $N_{\rm H_2}$ is the H$_2$ column density. 
The dust opacity  \citep{Ossenkopf1994} is
\begin{equation}
     K_\nu = 4.0({ \nu/505\,{\rm GHz} })^\beta \,{\rm cm}^2{\rm g}^{-1},
\end{equation}
and the dust emissivity index $\beta$ was fixed to 1.75 in the fitting \citep{Wang2015}. 

The derived $\rm H_2$ column density ($N_{\rm H_2}$) and dust temperature are shown in Fig.\,\ref{fig3}. $N_{\rm H_2}$ varies from $1.0\times 10^{20}$ to $1.4\times 10^{22}$\,cm$^{-2}$ with a mean value of $4.0\times 
10^{21}$\,cm$^{-2}$. 
The densest structure was identified as filament G323.179+0.149 by \citet{Li2016}. We plot its skeleton as the dot-dashed white line in Fig \ref{fig3}. 
Twenty-one highly reliable clumps from the Herschel Hi-GAL clump catalog \citep{Elia2017} and three clumps from the 
ATLASGAL dense clump catalog \citep{Urquhart2018} are located in the region we studied (their evolutionary stages 
and parameters are listed in table.\,\ref{table:1}). 
Because HIGALBM323.1050+0.0241 and AGAL323.104+0.024, HIGALBM323.1923+0.1539 and AGAL323.192+0.154 are 
the identical sources, there are a total of 22 clumps in the region we studied.

The dust temperature of the \object{G323.18+0.15} complex is relatively low (right panel of Fig.\,\ref{fig3}). This is especially true for the 
northwestern part. 
The H$\rm _{II}$ region G323.187+00.129, located between the northwestern and southeastern part, which was identified 
by \citet{Anderson2014}, has a higher temperature.

  \begin{figure*}
  \centering
  \includegraphics[trim={0cm 0 0 15.5cm},clip,width=18cm]{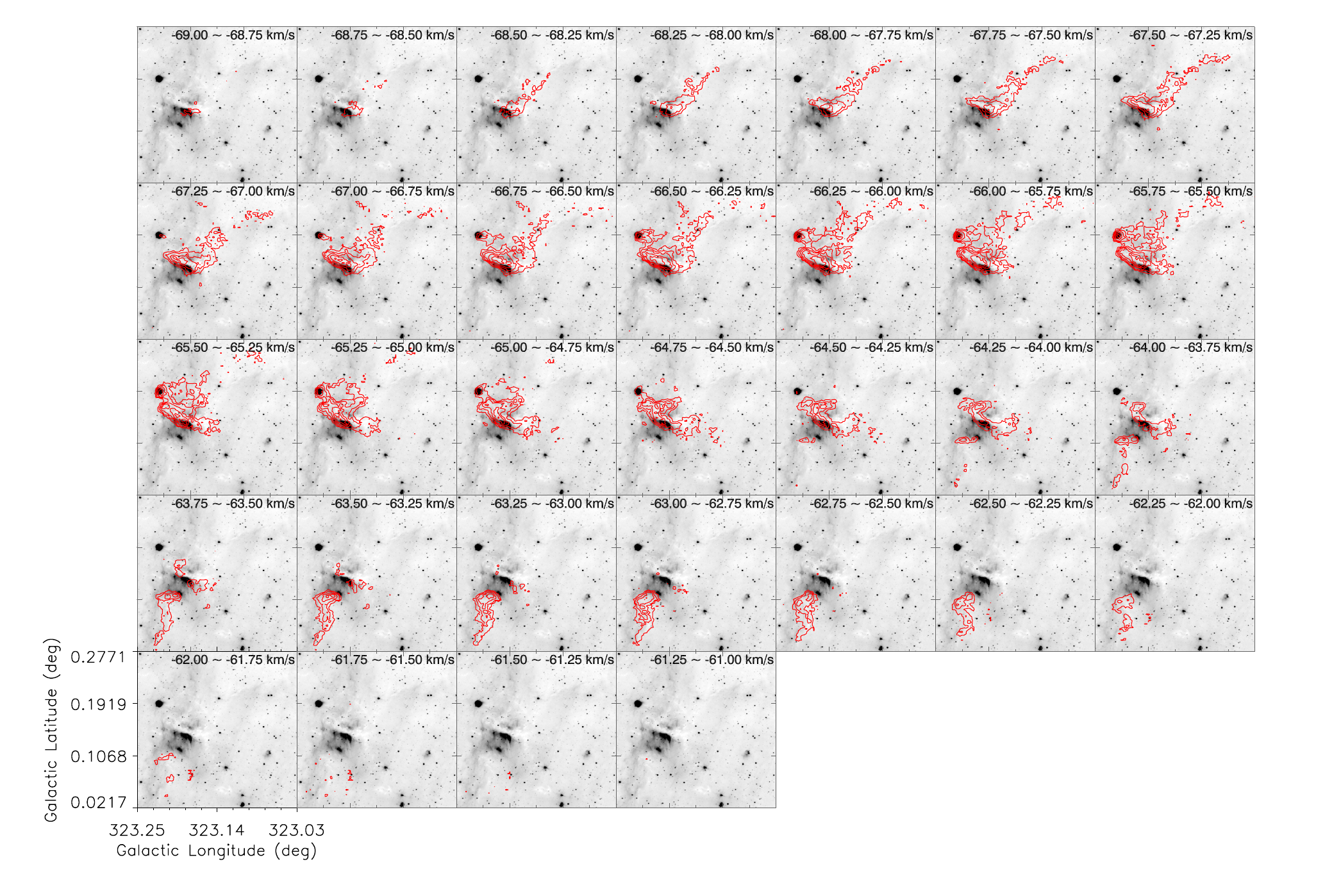}
  \caption{ Channel maps of the $^{13}$CO\,($J$\,=\,2--1) emission for \object{G323.18+0.15} complex (red contours) with contour levels at  $3$, $5$, $7$, $9$, and 11\,K\,km\,s$^{-1}$.
  The background is the Spitzer $8\,\mu$m emission map.}
    \label{fig4}
    \end{figure*}

   \begin{figure*}
   \centering
   \includegraphics[trim={0cm 3.0cm 0cm 6cm},clip, width=18cm]{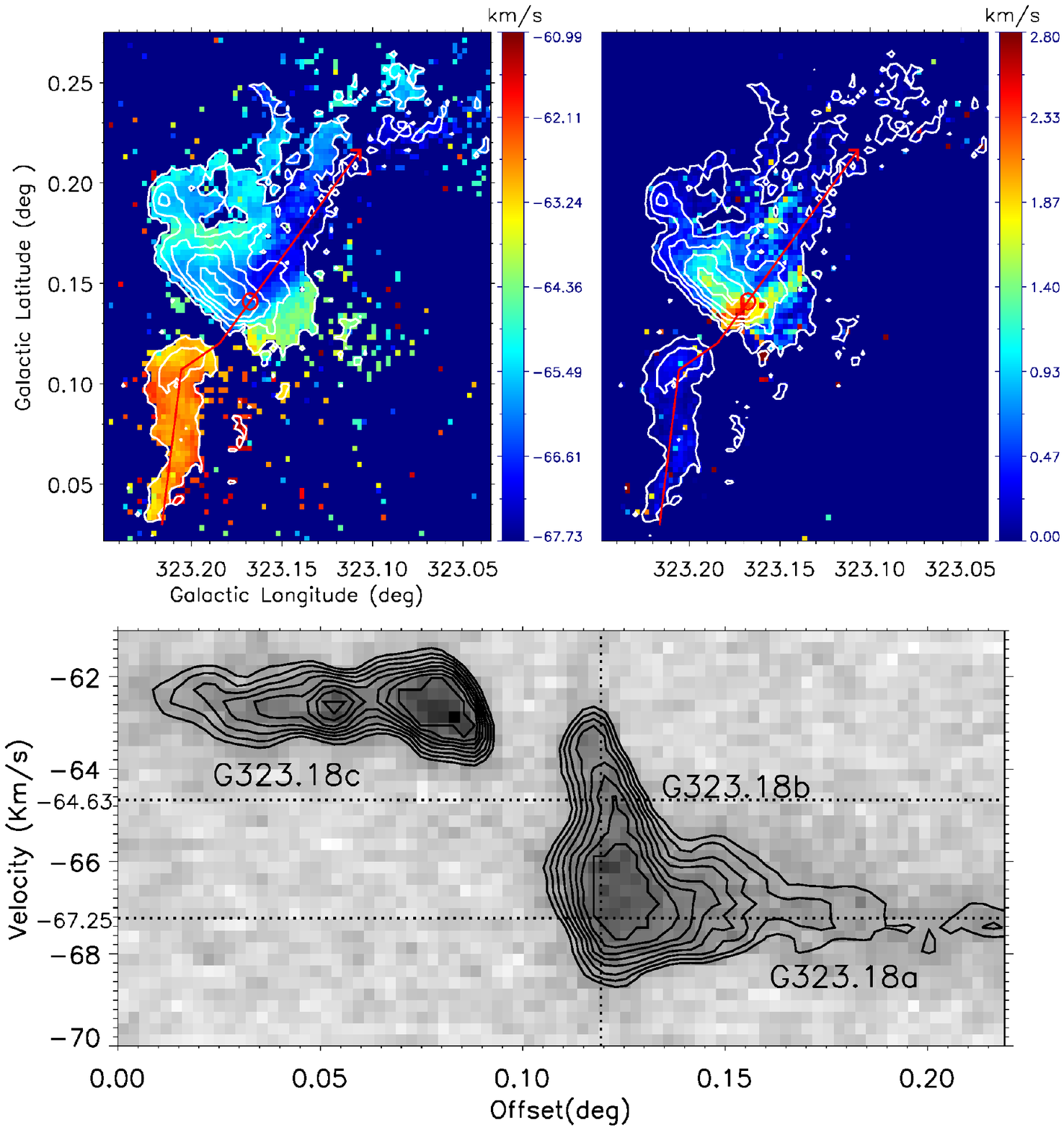}
      \caption{Moment\,1 and Moment\,2 of the \object{G323.18+0.15} complex, and PV diagram along its filamentary structure. Top: The color backgrounds in the left and right panels are velocity field and velocity dispersion of $^{13}$CO\,($J$\,=\,2--1). The white contours denote $^{13}$CO emission integrated from -69 to -61\,km\,s$^{-1}$, which starts from 2\,K\,km\,s$^{-1}$ (signal weaker than 3$\sigma$ is masked out) and increases in steps of 9\,K\,km\,s$^{-1}$.
      Bottom: Position-velocity diagram of the $^{13}$CO\,($J$\,=\,2--1) emission along the direction labeled by the red line in the top panel. The vertical dotted black line represents the position that is plotted as small red circles in the top panel. The horizontal  dotted black lines represent the system velocities of G323.18a and G323.18b. The black contours start at 3\,K and increase in steps of 0.9\,K.}
         \label{fig5}
   \end{figure*}

\subsection{Structure and kinematics}

The $^{13}$CO\,($J$\,=\,2--1) velocity channel maps in Fig.\,\ref{fig4} suggest the presence of three cloud components in the \object{G323.18+0.15} complex with different velocities: G323.18a with a velocity range from  $-69.0$ to $-65.5$\,km\,s$^{-1}$, the U-shaped structure G323.18b with a velocity range from $-65.5$ to $-63.75$\,km\,s$^{-1}$, and G323.18c with a velocity range from $-64.0$ to $-62.0$\,km\,s$^{-1}$. The system velocities of G323.18a, G323.18b, and G323.18c are estimated to be -67.25, -64.63, and -63\,km\,s$^{-1}$, respectively. The intensity-weighted velocity field of the $^{13}$CO\,($J$\,=\,2--1) emission (top left panel of Fig.\,\ref{fig5}) also shows the same three components.

\begin{figure}
  \centering
  \includegraphics[trim={2cm 0cm 0cm 1cm},clip,width=9cm]{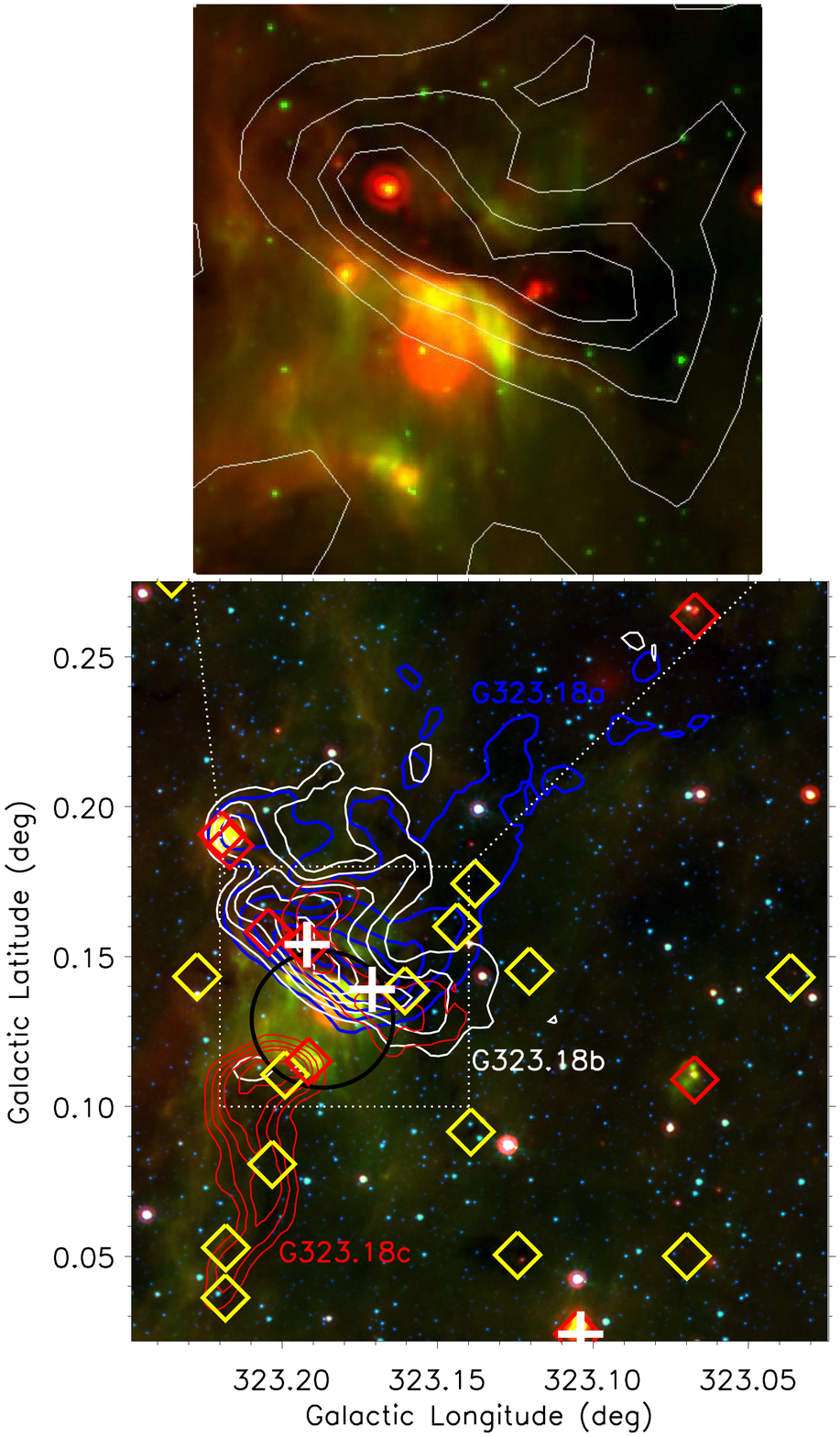}
  \caption{ Velocity-integrated intensity maps of the $^{13}$CO\,($J$\,=\,2--1) emission. 
  Bottom panel: Velocity-integrated intensity map of G323.18a (blue contours integrated from $-69.0$ to $-65.5$\,km\,s$^{-1}$ 
  starting at 2\,K\,km\,s$^{-1}$ with an increment of 5.5\,K\,km\,s$^{-1}$), 
  G323.18b (white contours, integrated from $-65.5$ to 64.0\,km\,s$^{-1}$ starting a 2\,K\,km\,s$^{-1}$ with an increment of
  3.8\,K\,km\,s$^{-1}$), 
  and G323.18c (red contours, integrated from $-64$ to $-62$\,km\,s$^{-1}$ starting at 2\,K\,km\,s$^{-1}$ with an increment of
  2.3\,K\,km\,s$^{-1}$). 
  The black circle denotes H$\rm_{II}$ region G323.187+00.129 identified by \citet{Anderson2014}. 
  The yellow and red diamonds are prestellar and protostellar clumps from the Herschel Hi-Gal clump catalog, and the white crosses 
  are ATLASGAL 870\,$\mu$m clumps. 
  The background is a three-color map with red, green, and blue for 24, 8, and 4.5\,$\mu$m, respectively. 
  Top panel: White contours start from 2\,$\times$\,10$^{21}$ cm$^{-2}$ and increase in steps of 1.9\,$\times$\,10$^{21}$ cm$^{-2}$. They display the distribution of the H$_2$ column density of the dotted rectangle in the bottom panel. 
  The background is the same as in the bottom panel.}
       \label{fig6}
  \end{figure}

Figure\,\ref{fig6} shows the velocity-integrated intensity map of $^{13}$CO\,($J$\,=\,2--1) for G323.18a, 
G323.18b, and G323.18c. It clearly shows the locations and structures of these three molecular cloud components. 
G323.18b appears to be a perfect U-shape structure with a compressed layer (the filament G323.179+0.149; \citet{Li2016})
with high-mass clumps in which high-mass stars are expected to form. 
The highest velocity dispersion also appears in the compressed layer, implying that there is strong turbulent activity 
(see the top right panel of Fig.\,\ref{fig5}). 
G323.18a is complementary to G323.18b. 
Some $^{13}$CO\,($J$\,=\,2--1) emission of G323.18c appears at the location of G323.18b, and
the H$\rm _{II}$ region G323.187+00.129 interacts with both G323.18b and G323.18c. This indicates that 
G323.18b and G323.18c are connected to each other. $^{12}$CO\,($J$\,=\,1--0) in Fig.\,\ref{fig1} also shows that they are in the same molecular cloud. 
As seen in the partially enlarged plot at the top of Fig.\,\ref{fig6}, the green 8\,$\mu$m emission seems to 
be blowing out at the southeastern side of the dark filament, and the red 24\,$\mu$m emission appears close to the 
southeastern side of the dark filament. 
High-mass stars appear to be forming in the compressed layer and contribute to exciting the H$\rm_{II}$ region G323.187+00.129 
\citep{Anderson2014}. 
Most of the strong 8 and 24\,$\mu$m emission seems to be wrapped inside G323.18b, which also supports this idea. 

We show the $^{13}$CO\,($J$\,=\,2--1) position-velocity (PV) diagram in the bottom panel of Fig. \ref{fig5}. 
The PV slice is extracted along the red line labeled in the top panels of Fig.\,\ref{fig5} and shows that G323.18c is separated from 
G323.18a and G323.18b, which may be because  H$\rm_{II}$ region G323.187+00.129 \citep{Anderson2014} blows off 
diffuse gas between them. 
The vertical dotted line along the offset axis corresponds to the position of the red circle on the red line in the top panels of Fig.\,\ref{fig5}, 
which identifies the compressed layer, that is, the interface between G323.18a and G323.18b. 
Strong CO emission appears between the system velocities of G323.18a and G323.18b, indicating that most of the CO from these 
two clouds is sufficiently mixed in the compressed layer. 
 
\begin{figure*}
   \centering
   \includegraphics[trim={0cm 0cm 0cm 14cm},clip,width=18cm]{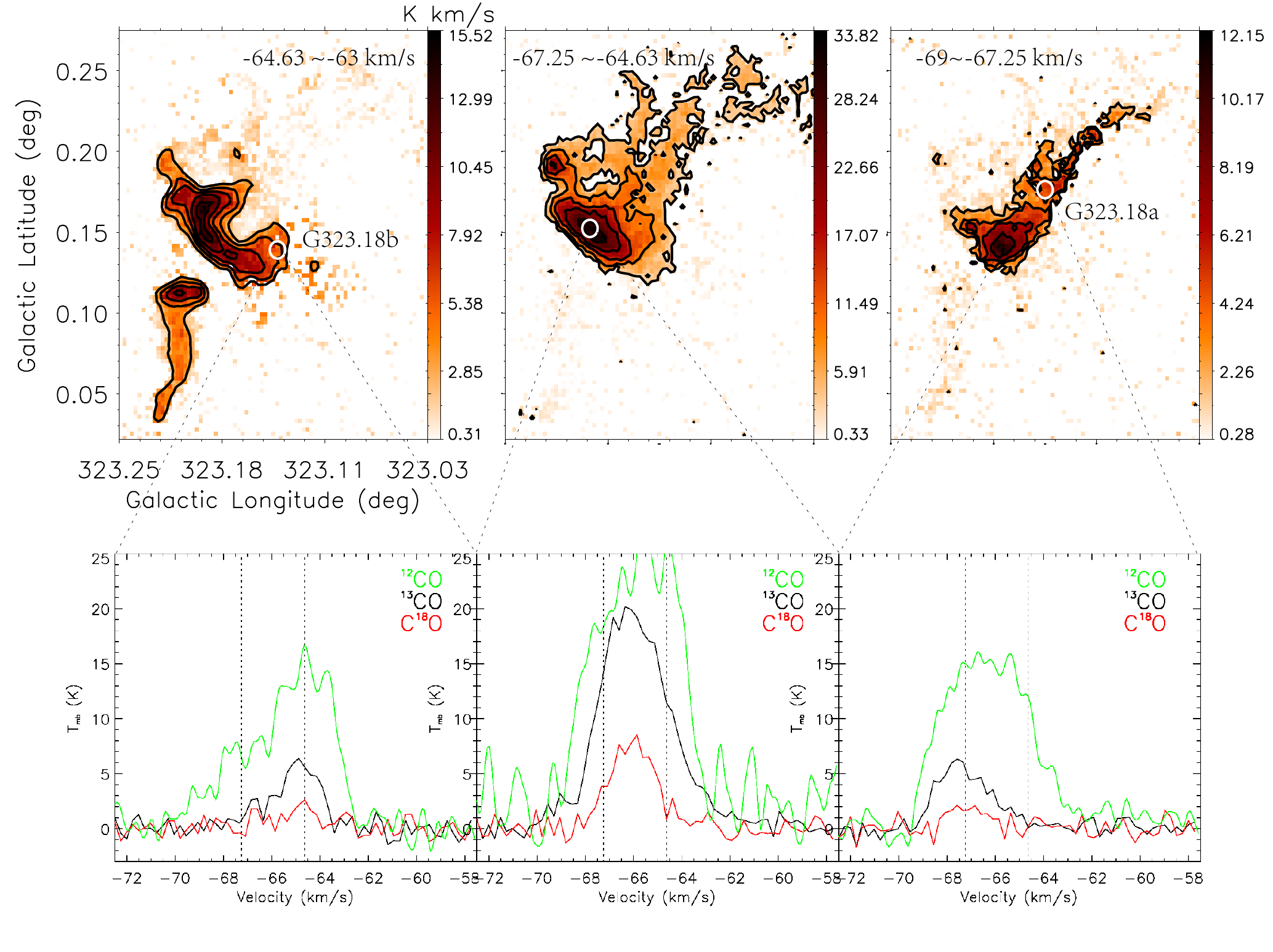}
  \caption{ Spectra of the three regions in \object{G323.18+0.15}. 
  The upper panels from left to right are the integral intensities corresponding to the velocity ranges from -64.63 to -63\,km\,s$^{-1}$, from 
  -67.25 to -64.63\,km\,s$^{-1}$, and from -69 to -67.25\,km\,s$^{-1}$. 
  The corresponding black contours start at 3\,K\,km\,s$^{-1}$ and increase in steps of 2.2, 6.1, and 1.3\,K\,km\,s$^{-1}$, respectively. 
  The lower panels are for the 3\,$\times$\,$^{12}$CO\,($J$\,=\,1--0), 1.5\,$\times$\,$^{13}$CO\,($J$\,=\,2--1), 
  and 1.5\,$\times$\,$C^{18}$O\,($J$\,=\,2--1) spectra extracted at three different positions of the \object{G323.18+0.15} complex, indicated 
  by white circles in the corresponding upper panels. 
  The vertical dashed black lines represent the systemic  velocities of G323.18a (-67.25\,km\,s$^{-1}$) and 
  G323018b (-64.63\,km\,s$^{-1}$).}
               \label{fig7}%
    \end{figure*}

 The upper panels of Fig.\,\ref{fig7}  show the integral intensity maps of $^{13}$CO\,($J$\,=\,2--1) corresponding to the 
 velocity ranges from -64.63 to -63\,km\,s$^{-1}$ (left panel), -67.25 to -64.63\,km\,s$^{-1}$ (middle panel), 
 and -69 to -67.25\,km\,s$^{-1}$ (right panel). 
 Their morphology indicates that the two clouds in the left and right panels may be interacting with each other, while the cloud in the 
 middle panel is the result of this interaction. 
 The lower panels show a 3\,$\times$\,$^{12}$CO\,($J$\,=\,1--0), 1.5\,$\times$\,$^{13}$CO\,($J$\,=\,2--1), and 
 1.5\,$\times$\,$C^{18}$O\,($J$\,=\,2--1) spectrum extracted at the positions indicated by the white circles in the corresponding 
 upper panels. 
 The $^{13}$CO and C$^{18}$O spectra in the right and left panels peak at the estimated system velocities of -67.25 (G323.18a) 
 and -64.63\,km\,s$^{-1}$ (G323.18b), while the $^{13}$CO and C$^{18}$O spectra in the middle panel peak at an 
 intermediate velocity between these two system velocities. 
 This is consistent with the above result, that is, the cloud in the middle panel is formed through an interaction between the two 
 clouds in the left and right panels, which causes the two initial clouds to appear as a single-peaked continuous cloud. 
 
 Two additional components lie at $\sim$ -52 and -30\,km\,s$^{-1}$, which are partly overlaid 
 on G323.18b in projection.
 Because they are very weak and their velocities are very different from that of the \object{G323.18+0.15} complex, we assume that these two 
 components just happened to be superimposed on the \object{G323.18+0.15} complex along the line of sight, and we do not show them here.

 \subsection{YSOs}
   The multiphase source classification scheme, a slightly modified version of the \citet{Gutermuth2009}, has been used to identify YSOs \citep{Zhou2020,Liu2015} (see Appendix A for details).  
   A total of 16 class\,I and 30 class\,II YSOs have been found in the region we studied (see Fig.\ref{fig2}). 
   Almost all class\,I YSOs are located in the U-shape structure G323.18b, while the class\,II YSOs around the U-shape structure 
   also display an arc-like distribution. 
   However, some class\,II YSOs may be associated with the H$\rm_{II}$ regions near the \object{G323.18+0.15} complex 
   (see Fig.\,\ref{fig2} and the discussion in sect.\,4.4). 


\section{Discussion}

    \subsection{Signatures of CCCs}

    First, we examined whether G323.18a and G323.18b are gravitationally bound as one structure. 
    The virial mass was determined using the expression \citep{Pillai2011,Issac2020}
      \begin{equation}
      M_{\rm vir} = \frac{5\sigma^2 R }{G} \,,
      \end{equation}
    where $R$ is the effective radius of the cloud and $\sigma\,=\,\Delta{V}/\sqrt{8\ln2}$ is the velocity dispersion. 
    Both $R$ and $\sigma$ were obtained using the $^{13}$CO\,($J$\,=\,2--1) line. 
    $R$ is thought of as ($A/ \pi)^{0.5}$, where $A$\,=\,26.46\,$\pm$\,5\,pc$^2$ is the area used to derive the cloud mass using the 
    area marked with the dotted white line in left panel of Fig.\,\ref{fig3}. This results in a value of $R$ = $2.9 \pm 0.31$\,pc. 
    From a Gaussian fit of the averaged $^{13}$CO spectrum of the cloud, we obtained the line width $\Delta$V \textasciitilde 3.6\,km\,s$^{-1}$. 
    This results in a value for the velocity dispersion of 1.5\,km\,s$^{-1}$. 
    Using these values, we computed the virial mass as $7.6 \,(\pm 0.8) \times 10^3$\,M$_\odot$. 
    The total gas mass of G323.18a and G323.18b was calculated using the expression
\begin{equation}
 M_{\rm H_2} = \mu m_{\rm H} \sum_{\rm i} S_{\rm i} \, N(\rm H_2)_{\rm i} \,,
\end{equation}
  where the mean molecular weight $\mu$ is 2.8, the mass of hydrogen atom is $1.67 \times 10^{-24}$\,g, 
  $S_{\rm i}$\,=\,(3.7\,$\pm$\,0.9)\,$\times 10^{35}$\,cm$^2$ is one pixel area, 
  and $N(\rm H_2)_{i\rm }$ is obtained from the spectral energy distribution fit. 
  We obtain a gas mass of 2.84\,($\pm$\,0.5)$\times10^3$\,M$_\odot$, which is lower than the virial mass. 
  This means that G323.18a and G323.18b as a whole cloud are not gravitational bound, which suggests that an accident 
  has triggered the ongoing star formation, for instance, a cloud-cloud collision.
We discuss the potential characteristic features of the CCC in the \object{G323.18+0.15} complex in the following subsections, including 
   the complementary distribution with displacement, the U-shape, the bridge and the fist moment distribution \citep{Fukui2018,Fukui2020}.

    \subsubsection{Complementary distribution and U-shape}


     
As shown in Fig.\,\ref{fig6}, G323.18b displays a perfect U-shape structure, and the cloud G323.18a complements the U-shape 
structure well. 
These two characteristic features suggest that G323.18a indeed collides with G323.18b. 
Based on the channel map shown in Fig.\,\ref{fig4}, we estimate the systemic velocities of G323.18a and G323.18b to be $-67.25$ 
and $-64.63$\,km\,s$^{-1}$. 
The actual collision velocity might be higher because of projection effects. 
As propounded by \citet{Issac2020}, \citet{Inoue2013}, and \citet{Fukui2015}, independent of the angle of two colliding clouds, the 
isotropic turbulence is enhanced at the collision-shocked layer. 
This is consistent with the fact that the velocity dispersion increases greatly along the U-shape structure (see the top right panel of Fig.\,\ref{fig5}).

\subsubsection{Bridge}
     Simulations of \citet{Haworth2015a,Haworth2015b} suggested that a broad intermediate-velocity feature that bridges between 
     two colliding clouds will appear in a PV diagram. 
     The bridge feature probes the turbulent motion of the gas enhanced by the collision and often appears at the spots of collisions
     \citep{Issac2020,Gong2017}. 
     The increased velocity dispersion along the U-shape, especially at the bottom where the compressed layer formed (see the top right 
     panel of Fig.\,\ref{fig5}), suggests that the compressed layer is the interface of the collision between G323.18a and G323.18b. 
     The bridge feature with an intermediate velocity between $-67.25$ and $-64.63$\,km\,s$^{-1} $ is clearly visible in the 
     PV diagram of \object{G323.18+0.15} in the bottom panel of Fig.\,\ref{fig5}. 
     The significant bridge feature indicates that G323.18a has interacted with G323.18b and that most of the gas of these two 
     clouds has been compressed into the dense layer at the bottom of U-shape. 
     This idea is also supported by the fact that both $^{13}$CO and C$^{18}$O spectrum from the compressed layer exhibit 
     a single-peak profile between the velocity $-67.25$ and $-64.63$\,km\,s$^{-1}$ (see Fig.\ref{fig7}). G323.18b and G323.18a may be in a later stage of CCC. They are very close to each other, and the bridge appears as the efficiently mixed gas of G323.18a and G323.18b.

\subsubsection{Distribution of the first moment}
        \citet{Fukui2018} showed simulation results of one small spherical cloud colliding with another large spherical cloud. 
        If the viewing angle between the line of sight and the direction of the collision is $0^{\circ}$ or $45^{\circ}$, the small cloud will 
        appear in the center of the large cloud in the first-moment map. 
        If the viewing angle is $90^{\circ}$, a compressed layer forms (see their Figures 4, 5, 6 and 7). 
        However, this may be not true when the colliding clouds are not spherical. 
        Both G323.18a and G323.18b appear in the first-moment map of $^{13}$CO\,($J$\,=\,2--1) (see the top left panel of Fig.\,\ref{fig5}), 
        and G323.18a  displays an extended filamentary structure that connects it to G323.18b. 
        This suggests that our viewing angle is not $0^{\circ}$ and that G323.18a is a filamentary cloud before it collides. 
        Considering that the velocity variation at the interface is relatively small (the bottom panel of Fig.\,\ref{fig5}) and because the compressed layer at the 
        interface between the two clouds appears as a filament, our viewing angle may be close to $90^{\circ}$, 

%
  \begin{figure*}
  \centering
  \includegraphics[width=15cm]{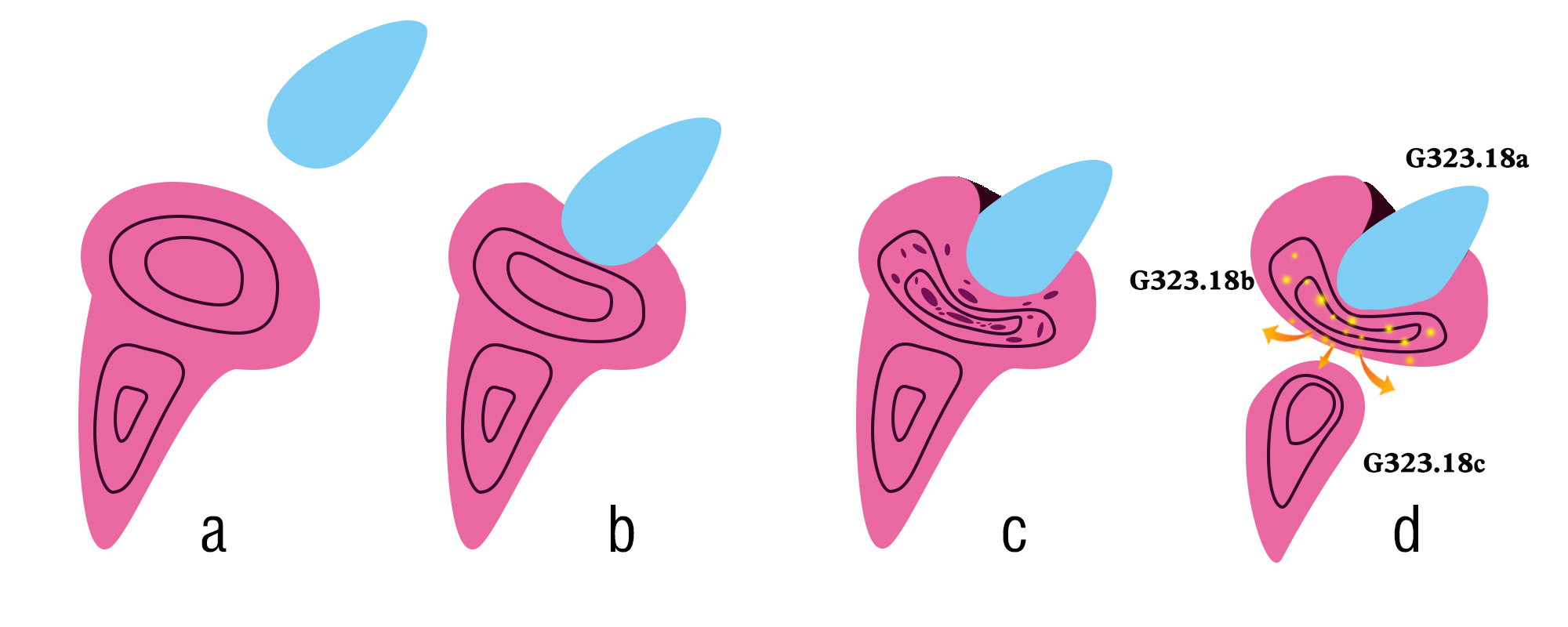}
      \caption{Schematic of the \object{G323.18+0.15} complex before and after the collision. 
      In panel a, the long blueshifted cloud G323.18a comes from the northwestern direction and collides with G323.18b and G323.18c. 
      In panels b and c, the collision forms a U-shape structure and a dense layer. 
      In panel d, YSOs are forming in the dense compressed layer, and then the strong stellar wind and radiation pressure 
      blow away the sparse gas and dust between G323.18b and G323.18c.
              }
        \label{fig8}
  \end{figure*}


\subsection{Induced filament, fragmentation of clumps, and star formation}

  Hydrodynamic simulations \citep{Habe1992,Anathpindika2010} suggest that the site of a CCC can be characterized by a 
  shock-compressed layer resulting from a bow shock that is driven by the smaller cloud into the larger cloud. 
  In the case of the \object{G323.18+0.15} complex, the shock-compressed layer manifests itself as an open U-shape structure. 
  This U-shape structure has a higher H$_2$ column density than the ambient medium, as shown in Fig.\,\ref{fig3}. 
  The compressed layer at the bottom of the U-shape structure is perpendicular to the direction of the collision.  

  The region we studied contains 22 clumps (see sect.\,3.2). 
  Thirteen of these 22 clumps are located in the \object{G323.18+0.15} complex (Figs.\,\ref{fig3} and \ref{fig6}), 
  that is, the area covered by $^{13}$CO\,($J$\,=\,2--1) emission with a signal-to-noise ratio S/N>3. 
  Their diameters and masses were recalculated using a distance of 3.55\,kpc. 
  The masses of these clumps range from 12.65 to 1.7 $\times10^{3}\,{\rm M_\odot}$ and the diameters range from 0.2 to 0.5\,pc. 
  Twenty-one clumps have surface densities above the threshold of 0.05\,g\,cm$^{-2}$ \citep{Urquhart2014} (see table.\,\ref{table:1}), and they
  are likely able to form high-mass stars.

  A total of 46 YSOs may been found in the whole region we studied using the GPSC (Fig.\,\ref{fig2}). 
  Eleven of the 16 class\,I YSOs and 15 of the 30 class\,II YSOs are located in \object{G323.18+0.15} and in the area covered by 
  $^{13}$CO\,($J$\,=\,2--1) emission with S/N>3. 
  Three dashed green circles are plotted in Fig.\,\ref{fig2}, where the areas of the circle and two annuli are the same as the area that is covered by 
  the molecular clouds G323.18a and G323.18b. 
  The average surface densities of class\,I and II YSOs in the innermost circle are 0.39\,arcmin$^{-2}$ and 0.58\,arcmin$^{-2}$, respectively,
  but they are 0.16 and 0.19\,arcmin$^{-2}$ in the first annulus, and 0 and 0.08\,arcmin$^{-2}$ in the second annulus. 
  It is evident that the surface density of YSOs in the interaction region is much higher. 
      
  Following \citet{Issac2020}, we used the full width at half maximum (FWHM) (3.56\,km\,s$^{-1}$) of the $^{13}$CO\,($J$\,=\,2--1) line extracted from the
  compressed layer to estimate the relative collision velocity, and calculated the collision timescale of cloud G323.18a and G323.18b. 
  With a cloud size 5.8\,pc, a collision timescale of $1.59 \pm 0.14$\,Myr is derived as an order-of-magnitude estimate. 
  This value may vary by a factor of 2 owing to projection effects in space and velocity and to the unknown configuration of 
  the clouds before the collision \citep{Fukui2014}. 
  For comparison, the ages of class I and II YSOs are 0.4 -- 0.7\,Myr and $2 \pm 1$\,Myr, respectively 
  \citep{Evans2009,Dunham2015,Issac2020}. 
  Our collision timescale of 1.59\,Myr exceeds the age of a class I YSO, but would be comparable to that of a class II YSO. 
  This suggests that most YSOs in the \object{G323.18+0.15} complex could have formed as a result of the collision between G323.18a and G323.18b, 
  especially for class I YSOs. 
  To some extent, this is consistent with the hypothesis that a collision triggered the YSOs formation.

 \subsection{CCC scenario for the \object{G323.18+0.15} complex}
   The images in Fig.\,\ref{fig8} presents a schema for the evolutionary development of the three cloud components 
   G323.18a, G323.18b, and G323.18c in the \object{G323.18+0.15} complex. 
   Our results suggest that G323.18b and G323.18c belong to the same filamentary cloud and that G323.18a collides with G323.18b. 
   In Fig.\,\ref{fig8}a, G323.18a comes from the northwestern direction and approaches the filamentary cloud structure, which 
   consists of G323.18b and G323.18c. 
   In Fig.\,\ref{fig8}b and Fig.\,\ref{fig8}c, G323.18a collides with G323.18bc and forms a U-shape structure and a compressed 
   dense layer at the bottom of G323.18b. 
   The colliding direction seems to be nearly perpendicular to our line of sight. 
   High-mass stars are forming within the dense layer, and together, they form an H$\rm _{II}$ region that provides strong feedback 
   that first breaks through the parent cloud from the southeastern side of the filament and heats and blows off diffuse gas between 
   G323.18b and G323.18c (Fig.\,\ref{fig8}d).

 \subsection{Possible effect from H$\rm _{II}$ regions}
  Although we find characteristic features of a CCC in \object{G323.18+0.15}, it is difficult to fully rule out other possibilities. 
  There are five H$\rm _{II}$ regions in the region we studied,  as identified by \citet{Anderson2014} (see Fig.\,\ref{fig2}). 
  In projection, the \object{G323.18+0.15} complex lies to the west of the largest H$\rm _{II}$ region G323.349+0.205 (traced by the largest white
  circle), and is close to the H$\rm _{II}$ region G323.101+0.268 (traced by the second largest white circle). 
  The \object{G323.18+0.15} complex seems to be on the rim of the 21\,cm continuum emission. However, the velocity distribution in the top right panel of Fig.\,\ref{fig5} does not show an increasing velocity gradient along the expanding direction of the H$\rm _{II}$ region (from east to west). Fig.\,\ref{fig2} shows that almost all class\,I YSOs lie along the U-shape structure G323.18b. Most class\,II YSOs are located in or nearby the region covered by G323.18ab. They do not lie along the left side of complex \object{G323.18+0.15}, where interaction may have taken place. This indicates that the largest H$\rm _{II}$ region may not strongly interact with complex \object{G323.18+0.15}, but we cannot rule out the possibility that some class\,II YSOs may be formed due to triggered star formation induced by the largest H$\rm _{II}$ region. The \object{G323.18+0.15} complex as a whole shows a filamentary structure that is nearly parallel to the expanding direction of the second largest H$\rm _{II}$ region (Fig.\,\ref{fig2}). G323.18a is closest to the H$\rm _{II}$ region and shows no evidence of being compressed by it, while G323.18b and G323.18c are far away from it. Therefore complex \object{G323.18+0.15} is probably not affected by the second largest H$\rm _{II}$ region.
  
  Due to the presence of the HII region in between filaments G323.18c and G323.18ab, we have to consider that the compression could be caused by the expansion of this H$\rm _{II}$ region into the gas. The H$\rm _{II}$ region G323.187+00.129 has a size of about 2.84\,pc. Our study of small bubbles with similar size, N22 \citep{Ji2012J} and N10 \citep{Ma2013}, suggests that the region may have a kinematic age younger than 1\,Myr. The collect-and-collapse model of triggered star formation may not work for the H$\rm _{II}$ region G323.187+00.129, that is, it has not had enough time to collect the surrounding materials and form compressed layers, and then to collapse to form clumps or cores and stars. As Fig.\,\ref{fig6} shows, G323.18ab covers nearly half of the H$\rm _{II}$ region in projection, but no compressed CO or dust shells appear along the edge of the H$\rm _{II}$ region. However, the northwestern part of G323.18c shows some evidence of being compressed by the H$\rm _{II}$ region. This is consistent with the idea we described in sect.\,4.3.
  
\section{Conclusions}
   The structure and kinematics of the \object{G323.18+0.15} complex has been studied, and strong evidence is found that supports the hypothesis 
   that a CCC occurred between the cloud components G323.18a and G323.18bc. 
   Our results are listed below.
   \begin{enumerate}
      \item The G323.18a - G323.18b structure as a whole is probably not gravitationally bound, suggesting that any star formation 
       in these clouds has been triggered by an accident. 
       A reasonable explanation is that the smaller cloud G323.18a collides with the larger cloud G323.18bc, which forms a 
       U-shape structure and a dense compressed layer in G323.18b. 
       The column density of the compressed layer is estimated at $1.36\times10^{22}$\,cm$^{-2}$, which satisfies the 
       threshold condition for forming high-mass stars. 

    \item G323.18b appears as a perfect U-shape structure showing an enhanced velocity dispersion. 
    The shape of the G323.18a structure complements the U-shape structure found in G323.18b. 
    Except for these two typical features of a CCC, the bridge feature with an intermediate velocity between G323.18a and 
    G323.18b also supports the hypothesis that they collide.

      \item A total of 46 YSOs were found to be distributed in the region we studied. 
      The average surface density of class\,I and class\,II YSOs in the region covered by G323.18a and G323.18b is 
      about 0.39 and 0.58\,arcmin$^{-2}$, respectively, which is much higher than for the surrounding regions.       
      The timescale of the collision of G323.18a and G323.18b is estimated to be 1.59\,Myr, which is longer than the 
      typical lifetime of class I YSOs and is comparable to the lifetime of class II YSOs. 
      Almost all class I YSOs are associated with the U-shape structure and suggest a CCC-related origin.
   \end{enumerate}

   \begin{table*}
\caption{ Parameters of dust cores associated with \object{G323.18+0.15}. 
   The first, second, and third columns list the names, diameters, and masses of the clumps in the Herschel Hi-Gal clump catalog. 
   The fourth and fifth columns indicate the evolutionary stage of the clumps and whether it is located within the \object{G323.18+0.15} complex. 
   The sixth and seventh columns indicate the new diameters and masses calculated for a distance of 3.55 kpc. 
   The eighth column presents the clump surface density. 
   AGAL323.104+0.024, AGAL323.192+0.154, and AGAL323.171+0.139 are the clumps that fall into the \object{G323.18+0.15} complex 
   in the ATLASGAL dense clump catalog. 
   The sources AGAL323.104+0.024, AGAL323.192+0.154, HIGALBM323.1050+0.0241, and HIGALBM323.1923+0.1539 are duplicated.}             

\label{table:1}      
\centering                          
\begin{tabular}{p{4cm}p{0.8cm}p{1.0cm}p{1.8cm}p{0.5cm}p{1.3cm}p{1.3cm}p{2.8cm}}
\hline\hline                 
Name & Diam & Mass & Evol & in/out & Diam(new) & Mass(new) &surface density  \\    
     & pc   & M$_\odot$ &   &   &pc & M$_\odot$ & g\,cm$^{-2}$\\
\hline                        

HIGALBM323.0365+0.1431 & 1.158 &   425.61 & prestellar  & out &       &        \\
HIGALBM323.0672+0.2635 & 0.471 &   409.01 & protostellar& out &       &       \\
HIGALBM323.0673+0.1090 & 1.270 &  1079.55 & protostellar& out &       &       \\
HIGALBM323.0698+0.0501 & 1.463 &  2190.30 & prestellar  & out &       &       \\
HIGALBM323.1050+0.0241 & 0.816 &   510.81 & prestellar  & out &       &       \\
(AGAL323.104+0.024) &                                              \\
HIGALBM323.1205+0.1452 & 1.151 &  1436.53 & prestellar  & out &       &       \\
HIGALBM323.1243+0.0505 & 1.885 &  5153.03 & prestellar  & out &       &       \\
HIGALBM323.1379+0.1743 & 0.481 &   307.05 & prestellar  & in  & 0.393 &204.97 & 0.35 \\
HIGALBM323.1393+0.0916 & 1.079 &  1063.05 & prestellar  & out &       &       \\
HIGALBM323.1437+0.1600 & 0.921 &  1771.21 & prestellar  & in  & 0.350 &256.37 & 0.56 \\
HIGALBM323.1605+0.1387 & 1.392 & 11753.03 & prestellar  & in  & 0.530 &1701.18 & 1.62\\
HIGALBM323.1915+0.1151 & 1.157 &   508.43 & protostellar & in  & 0.439 & 73.28 & 0.10 \\
HIGALBM323.1923+0.1539 & 0.614 &   535.31 & protostellar & in  & 0.233 & 77.15 & 0.38 \\
(AGAL323.192+0.154)                             \\
HIGALBM323.1990+0.1106 & 0.916 &  1483.38 & protostellar & in & 0.348 &213.79 & 0.47 \\
HIGALBM323.2032+0.0808 & 1.538 &  1350.40 & prestellar  & in & 0.464 &122.81 & 0.15\\
HIGALBM323.2044+0.1582 & 0.530 &   314.52 & protostellar& in & 0.201 &45.33 & 0.30 \\
HIGALBM323.2168+0.1870 & 1.567 &   134.32 & protostellar& in & 0.481 &12.65 & 0.01 \\
HIGALBM323.2181+0.0529 & 1.697 &  1102.78 & prestellar  & in & 0.512 &100.29 & 0.10\\
HIGALBM323.2182+0.0362 & 1.445 &  1626.27 & prestellar  & in & 0.436 &147.89 & 0.21 \\
HIGALBM323.2194+0.1907 & 1.479 &  1014.20 & protostellar& in & 0.454 &95.50 & 0.12 \\
HIGALBM323.2273+0.1433 & 1.119 &  3038.81 & prestellar  & out&       &       \\
AGAL323.171+0.139      &0.468 & 749.89 & YSO  &  in & 0.413 &584.79 & 0.92\\

\hline                                   
\end{tabular}
\end{table*}

\begin{acknowledgements}
     The authors thank the anonymous referee for helpful comments. This work was mainly funded by the National Natural Science foundation of China (NSFC) under grant No.11973076. It was also partially supported by the NSFC under grant Nos.11433008, 11903070, 12173075, and 12103082, the Heaven Lake Hundred Talent Program of Xinjiang Uyghur Autonomous Region of China, the Natural Science Foundation of Xinjiang Uygur Autonomous Region under grant No. 2022D01E06, and the CAS Light of West China Program under grant Nos. 2020-XBQNXZ-017 and 2021-XBQNXZ-028. D. L. Li has been funded by Youth Innovation Promotion Association CAS and Tianshan Innovation Team Plan of Xinjiang Uygur Autonomous Region (2022D14020).
     Y. X. has been funded by High Level Talent Heaven Lake Program of Xinjiang Uyghur Autonomous Region of China. W. B. has been funded by Chinese Academy of Sciences President’s International Fellowship Initiative by Grant No. 2021VMA0008.
     
      
      
\end{acknowledgements}

\bibliographystyle{aa}
\bibliography{ref}

\begin{appendix}
\section{Young star classification}

      First, we selected the candidates that are located in the \object{G323.18+0.15} complex and have photometric detections in all four 
      IRAC bands ($\mathrm{\sigma < 0.2\, mag}$) from the GLIMPSE I Spring 07 highly reliable catalog. 
      Then we removed sources contaminated by polyciclic aromatic hydrocarbons according to the constraints from \citet{Gutermuth2009}: 
      $\mathrm{[3.6]-[4.5] - \sigma \leq 1.4 \times (([4.5]-[5.8]) + \sigma_1 - 0.7) + 0.15}$ and $\mathrm{[3.6]-[4.5] - \sigma_2 < 1.65}$, 
      where $\mathrm{\sigma_1 =\sigma \, {\left\{[4.5]-[5.8]\right\}}}$ and $\mathrm{\sigma_2 =\sigma \, {\left\{[3.6]-[4.5]\right\}}}$. 
      Finally, we identified 16 class I YSOs following the constraints [4.5]-[5.8] > 0.7 and [3.6]-[4.5] > 0.7, and 12 class II YSOs 
      following the constraints $\mathrm{[4.5] - [8.0] - \sigma_3 >0.5}$, $\mathrm{[3.6] - [5.8] - \sigma_4 >0.35}$, 
      $\mathrm{[3.6] - [5.8] + \sigma_4 \leq 3.5 \times (([4.5] - [8.0]- \sigma_3) -0.5) +0.5}$, 
      and $\mathrm{[3.6]-[4.5] - \sigma_2 > 0.15}$, where $\mathrm{\sigma_3 =\sigma \, {\left\{[4.5]-[8.0]\right\}}}$ and 
      $\mathrm{\sigma_4 =\sigma \, {\left\{[3.6]-[5.8]\right\}}}$. 
      Four sources lack detections at either 5.8 or 8.0\,$\mu$m but with 3.6 and 4.5\,$\mu$m with $\mathrm{\sigma < 0.2\, mag}$ 
      and H and K bands from 2MASS with $\mathrm{\sigma < 0.1\, mag}$ in our region. 
      We identified 24 class II YSOs following the constraints $\mathrm{[[3.6] - [4.5]]_0 - \sigma_1 >0.101}$, 
      $\mathrm{[K - [3.6]]_0 - \sigma_2 >0}$, 
      $\mathrm{[K - [3.6]]_0 - \sigma_2 > -2.85714 \times ([[3.6]- [4.5]]_0 - \sigma_1 - 0.101)\,}$ + 0.5 , and $[3.6]_0 < 14.5$ 
      where $\mathrm{\sigma_1 = \sigma \,{\left\{[3.6] - [4.5]\right\}_{meas}}}$ 
      and $\mathrm{\sigma_2 = \sigma \, {\left\{[K] - [3.6]\right\}_{meas}}}$ \citep{Gutermuth2009}. 
      We used MIPS $\mathrm{24 \mu m}$ with magnitude [24] < 7 ($\sigma$ < 0.2 mag) to identify transition disks (TDs) 
      following the criteria [5.8]-24] > 2.5 or [4.5]-24] > 2.5, and [3.6] < 14. 
      The TDs are classified as sources in an evolutionary stage between class II and class III \citep{Gutermuth2009}, but no TD was found in our \object{G323.18+0.15} complex. 
      The absence of a TD suggests that star formation in \object{G323.18+0.15} is in its early stage, and perhaps the CCC between 
      G32.18a and G323.18b led to a rapid increase of mass at the interface of the collision, and was followed by star formation in a short time.
      
\end{appendix}

\end{document}